\documentclass{aa}
\usepackage{graphicx}
\begin{document}
   \title{A photometric study of the young open cluster
NGC~1220\thanks{Based on observations
carried out at Observatorio de S. Pedro Martir, 
UNAM, Mexico},\thanks{Photometry is only available
in electronic form at the CDS via anonymous ftp to {\tt cdsarc.u-strasbg.fr
(130.79.128.5} or via {\tt http://cdsweb.u-strasbg.fr/cgi-bin/qcat?J/A+A//}}}
\author{
S. Ortolani\inst{1}
\and
G. Carraro\inst{1}
\and
S. Covino\inst{2}
\and
E. Bica\inst{3}
\and
B. Barbuy\inst{4}
}
\institute{
Dipartimento di Astronomia, Universit\`a di Padova,
Vicolo Osservatorio 2, I-35122 Padova, Italy\\
e-mail: ortolani@pd.astro.it, giovanni.carraro@unipd.it
\and
Osservatorio Astronomico di Milano, Italy\\
e-mail: covino@merate.mi.astro.it
\and 
Universidade Federal do Rio Grande do Sul, Dept. de
Astronomia, CP 15051, 
                   Porto Alegre 91501-970, Brazil\\
e-mail:bica@if.ufrgs.br
\and
Universidade de Sao Paulo, Dept. de Astronomia, CP 3386, Sao
Paulo 01060-970, Brazil\\
e-mail:barbuy@astro.iag.usp.br
            }
\date{Received March 2002; accepted}
\abstract{ We present UBV  CCD observations obtained in the field
of the northern open cluster NGC\,1220, for which little information
is available.
We provide also BV CCD photometry of a field 5$^{\prime}$ northward
of NGC~1220 to take into account field star contamination.
We argue that NGC~1220 is a young compact open cluster, for
which we estimate a core radius in the range $1.5-2.0$ arcmin.
We identify 26  likely candidate members 
with spectral type earlier than $A5$, down to $V_o$=15.00 mag 
on the basis of the position in the two-colour Diagram
and in the Colour Magnitude Diagrams (CMDs).
By analyzing the distribution of these stars in the colour-colour
and CMDs, we find that
NGC\,1220
has a reddening E$(B-V)=0.70\pm0.15$ mag, is placed 
$1800\pm200$ pc distant from the Sun, and has an age 
of about 60 Myrs. The cluster turns out to be located
about 120 pc above the Galactic plane, relatively high with respect 
to its age. 
 }
\titlerunning{Photometry of NGC~1220}
\authorrunning{S. Ortolani et al.}
\maketitle
%

\section{Introduction}
NGC\,1220 (Collinder~37, OCL~380) is an open cluster in Perseus for which not much information is
available. 
Its near anticenter projection (l = 143.03$^{\circ}$, 
b= -3.96$^{\circ}$; J2000 $\alpha$ = 3$^{\rm h}$11$^{\rm m}$40$^{\rm s}$, 
$\delta$ = 53$^{\circ}$20'45''), small angular size ($\approx$ 3') and 
its contrasted appearance with respect to the
field, as inspected on DSS (Digitized Sky Survey) images, make it
an interesting target.
To our knowledge, this cluster remained unstudied insofar, but
for the very preliminary investigation by Phelps et al. (1994),
who presented BVI instrumental CMDs.
Since they were looking for old open cluster candidates, their
analysis was limited to the remark that this cluster has to be young,
since no clear giant branch and clump were detected.
Therefore, these authors do not report any estimates of the cluster
fundamental parameters.
In the present study we provide the first 
calibrated UBV CCD photometry of NGC\,1220, aiming at determining its basic parameters, such as reddening, distance and age, which are fundamental
to understand the disk sub-system which the cluster belongs to.

\noindent
The plan of this paper is as follows.\\
In Sect.~2 we present the observations and data reduction and
in Sect.~3 we derive an estimate of the cluster diameter.
In Sect.~4 we describe the CMDs,
whilst in Sect.~5 we derive cluster reddening, and isolate
candidate members.
Sect.~6 is dedicated to derive NGC~1220 distance and age.
Finally, in Sect.~7 we draw some
conclusions and suggest further lines
of research.

   \begin{figure}
   \centering
   \resizebox{\hsize}{!}{\includegraphics{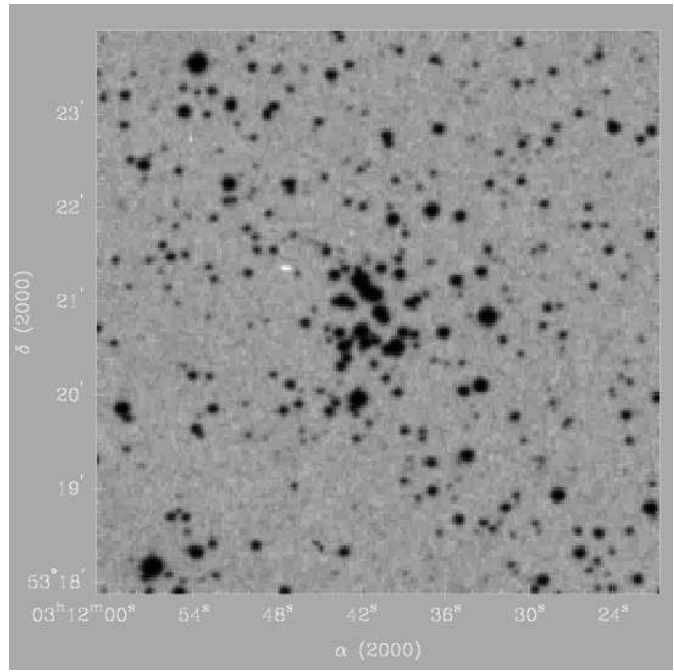}} 
   \caption{A DSS red map of the covered region in the field
of NGC~1220. North is up, East on the left.}
    \end{figure}
   \begin{figure}
   \centering

   \resizebox{\hsize}{!}{\includegraphics{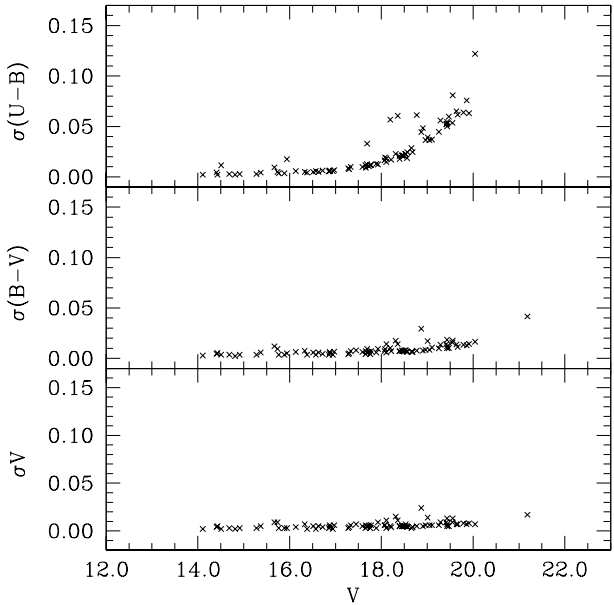}} 
   \caption{Run of the magnitude and colours errors as a function of V for our
observations of NGC~1220}
    \end{figure}

\section{Observations and Data Reduction}
Observations were carried out with the CCD  SIT $\#1$ camera at the 
Observatorio Astron\'omico Nacional (OAN) de S. Pedro Martir, B.C.,
Mexico, in the photometric
nights of September 16-17, 
2001. This CCD samples a $4^\prime.6\times4^\prime.6$ field in a  
$1K\times 1K$ thinned CCD. The typical seeing was around 1.7 
arcsec.  
 
The details of the observations are listed in Table~1 and Table~2,
where the observed fields are reported together with the exposure
times, the typical seeing and the airmass from the night of September 16 and
17, respectively.
The covered region is shown in Figs.~1, where a DSS\footnote
{Digital Sky Survey, {\tt http://archive.eso.org/dss/dss}} 
map is  presented for NGC~1220.

The data have been reduced by using the MIDAS and DAOPHOT packages.
The calibration equations obtained by observing Landolt (1992) 
SA~92, SA~95, PG~0220+015, PG~2331+055 and PG~0231+051,PG~2336+004 and
PG~1633+099 fields along both nights for a total
of 58 independent measurements, , are: 
	\begin{eqnarray}  
\nonumber 
u \! &=& \! U + 19.56\pm0.015 + (0.04\pm0.02)(U\!-\!B) + 0.42\,X \\  
\nonumber 
b \! &=& \! B + 21.23\pm0.015 - (0.08\pm0.017)(B\!-\!V) + 0.22\,X \\  
\nonumber 
v \! &=& \! V + 21.90\pm0.015 + (0.015\pm0.010)(B\!-\!V) + 0.14\,X \\  
	\label{eq_calib} 
	\end{eqnarray} 
where $UBV$ are standard magnitudes, $ubv$ are the instrumental  
ones, and $X$ is the airmass. The zero points are for 1 sec exposure time.
The standard stars in these fields
provide a very good colour coverage from $-0.33 \leq (B-V) \leq 1.44$.
Only one standard star has been excluded because of its large deviation
from the average value ($SA~95-106$).
For the extinction coefficients, 
we assumed the typical values of the site of 0.42, 0.22 and 0.14 fro $U$, $B$
and $V$ passbands, respectively, available from the OAN web site
\footnote{http://bufadora.astrosen.unam.mx}.
The photometric errors are
presented in Fig.~2 for $V$, $(B-V)$ and $(U-B)$.
The zero point errors of the 
final photometry should include also the transfer from the 
DAOPHOT convolved magnitudes to the wide aperture ones 
(6.7 arcsecs radius) needed to avoid any PSF and seeing variation effect. This transfer accounts for an additional $\pm$0.02 mag uncertainty 
on all colours.

  \begin{figure}
   \centering
   \resizebox{\hsize}{!}{\includegraphics{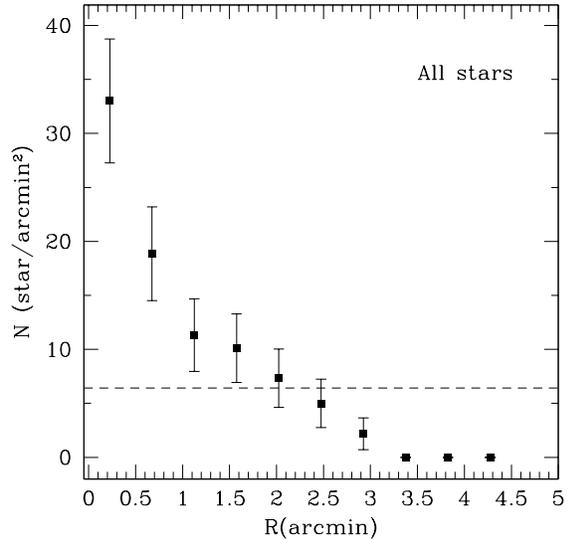}} 
   \caption{Star counts in the field of of NGC~1220 as a function of
   radius. The dashed line indicates the background level, as derived 
   from star counts in the accompanying field.}
    \end{figure}

\begin{table} 
\tabcolsep 0.15truecm 
\caption{Journal of observations of NGC~1220, 
the accompanying field, and standard star fields (September 16, 2001).} 
\begin{tabular}{ccccc} 
\hline 
\multicolumn{1}{c}{Field}    & 
\multicolumn{1}{c}{Filter}    & 
\multicolumn{1}{c}{Time integration}& 
\multicolumn{1}{c}{Seeing}       &
\multicolumn{1}{c}{Airmass} \\
      &        & (sec)     & ($\prime\prime$)&\\ 
  
\hline 
 SA~95          &     &              &      &     \\
                & $U$ &  120         &  1.7 & 1.21\\
                & $B$ &  45          &  1.8 & 1.20\\ 
                & $V$ &  30,30       &  1.6 & 1.22\\ 
 NGC~1220       &     &              &      &     \\ 
                & $U$ &  120         &  1.8 & 1.16\\ 
                & $B$ &  120,1800    &  1.9 & 1.13\\ 
                & $V$ &  30,900      &  1.6 & 1.14\\ 
PG~1633+099     &     &              &      &     \\ 
                & $U$ &  90          &  1.9 & 1.38\\
                & $B$ &  40          &  1.8 & 1.37\\ 
                & $V$ &  20          &  1.7 & 1.39\\ 
PG~0231+051     &     &              &      &     \\ 
                & $U$ &  90          &  1.6 & 1.11\\
                & $B$ &  30,30       &  1.7 & 1.12\\ 
                & $V$ &  20,20,20    &  1.7 & 1.12\\ 
PG~0220+015     &     &              &      &     \\ 
                & $U$ &  90          &  1.6 & 1.12\\
                & $B$ &  30          &  1.7 & 1.12\\ 
                & $V$ &  20,20       &  1.8 & 1.12\\ 
\hline 
\end{tabular} 
\end{table} 

\begin{table} 
\tabcolsep 0.15truecm 
\caption{Journal of observations of NGC~1220, 
the accompanying field, and standard star fields (September 17, 2001).} 
\begin{tabular}{ccccc} 
\hline 
\multicolumn{1}{c}{Field}    & 
\multicolumn{1}{c}{Filter}    & 
\multicolumn{1}{c}{Time integration}& 
\multicolumn{1}{c}{Seeing}       &
\multicolumn{1}{c}{Airmass} \\
      &        & (sec)     & ($\prime\prime$)&\\ 
  
\hline 
 SA~92          &     &              &      &     \\
                & $U$ &  90          &  1.7 & 1.16\\
                & $B$ &  30          &  1.8 & 1.16\\ 
                & $V$ &  20,20       &  1.6 & 1.16\\ 
 NGC~1220       &     &              &      &     \\ 
                & $U$ &  1200        &  1.6 & 1.13\\ 
PG~2336+004     &     &              &      &     \\ 
                & $U$ &  90          &  1.9 & 1.20\\
                & $B$ &  30          &  1.8 & 1.21\\ 
                & $V$ &  20,20       &  1.7 & 1.21\\ 
PG~0231+051     &     &              &      &     \\ 
                & $U$ &  90          &  1.6 & 1.13\\
                & $B$ &  30,30       &  1.7 & 1.13\\ 
                & $V$ &  20,20       &  1.7 & 1.14\\
Field           &     &              &      &     \\
                & $B$ &  1200        &  1.6 & 1.09\\ 
                & $V$ &  20,600      &  1.6 & 1.10\\ 
PG~2331+055     &     &              &      &     \\ 
                & $U$ &  90          &  1.6 & 1.11\\
                & $B$ &  30,30       &  1.7 & 1.11\\ 
                & $V$ &  20,20       &  1.8 & 1.14\\ 
\hline 
\end{tabular} 
\end{table}

   \begin{figure*}
   \centering
   \resizebox{\hsize}{!}{\includegraphics{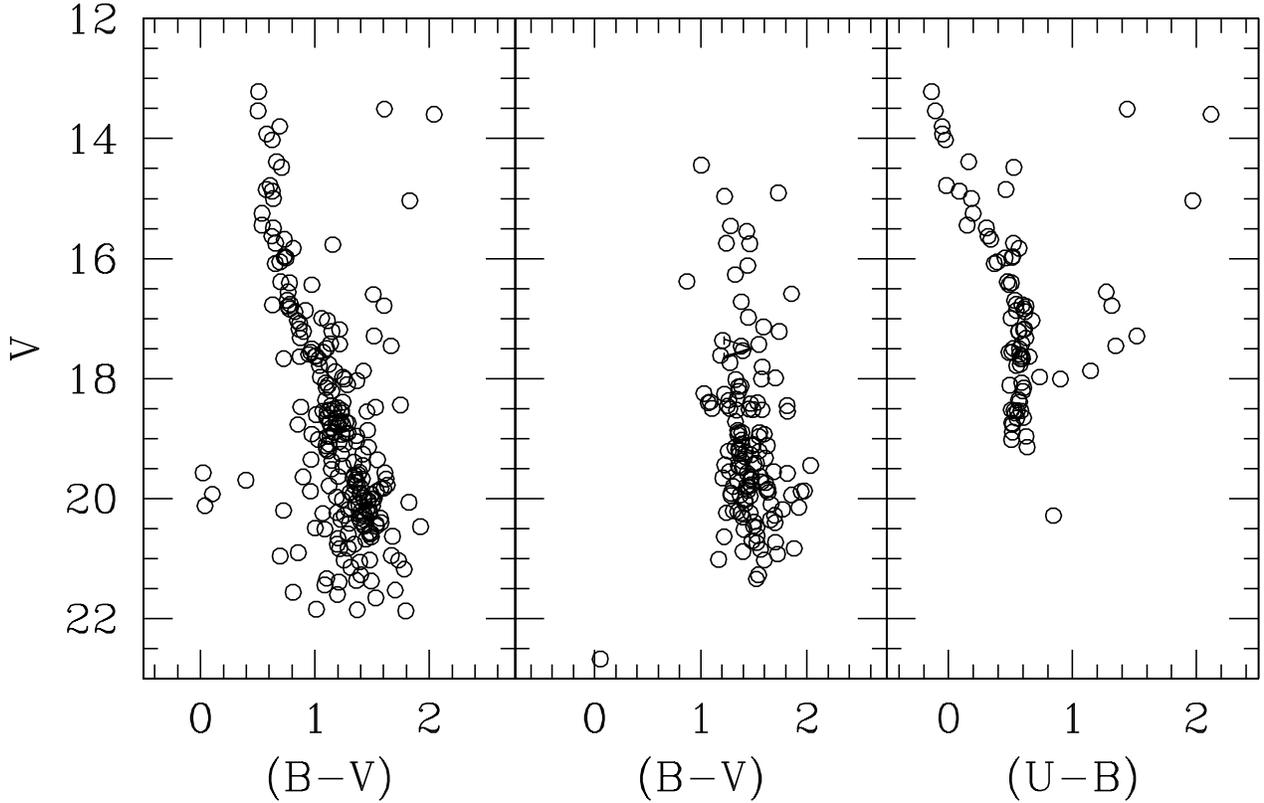}} 
   \caption{ CMDs of the stars in the region of NGC~1220. {\bf
   Left panel}:
    all NGC~1220 stars in the $V$ vs $(B-V)$ plane. {\bf Middle panel}:
all the stars in the accompanying field  in the $V$ vs $(B-V)$ plane.{\bf
   Right panel}:
    all NGC~1220 stars in the $V$ vs $(U-B)$ plane}
    \end{figure*}

\section{Star counts and cluster size}
NGC~1220 appears as a compact cluster of about 20 stars (see Fig.~1).
According to Lyng\aa~ (1987), NGC~1220 has a diameter of 2 arcmin,
so our study covers the entire cluster region.
To infer an estimate of the radius,
we derive  the surface stellar density by performing star counts
in concentric rings around the center of the covered area,
and then dividing by their
respective surface areas. The final density profile and the corresponding
Poisson error bars are depicted in Fig.~3.
The dashed line in this plot represents the star counts in the accompanying
field,
where we estimated 6.3 {\rm stars/arcmin$^{2}$} .
The cluster clearly dominates over the field up to about 2 arcmin,
then it completely merges with the Galactic disk field star component.
Therefore, we estimate a cluster radius of about 1.5-2.0 arcmin, 
somewhat larger than the value reported by Lyng\aa~(1987).

\section{Colour-Magnitude Diagram}
The CMD for all the stars measured in the
direction of NGC~1220 is shown in Fig.~4. In the left
panel we plot all the stars in the $V$ vs $(B-V)$ plane,
where in the middle panel we plot in the same plane
stars from the accompanying field observed $5^{\prime}$
northward of NGC~1220.
Finally, in the left panel we present the CMD in the plane
$V$ vs $(U-B)$.
These CMDs are easy to interpret. On the left panel, 
the cluster Main Sequence (MS) extends
almost vertically from $V \approx 13$ up to $V \approx 21$,
although it starts widening at $V \approx 19$.
This is clearly related to the field star component,
which is heavy in the direction of NGC~1220, a cluster
located quite low in the Galactic plane. 
In fact (see the middle panel) the contribution
of field stars is quite heavy in the magnitude interval
$18 \leq V \leq 21$ mag, and some stars on the red side
of the MS most probably belong to the field.
The appearance of this CMD suggests that we are facing a young
cluster.
The three very red stars in the left panel might be the signature
of a possible evolved population in NGC\,1220.
We claim that this is very difficult. First of all,
the faintest star of the triplet has a nice counterpart in the middle
panel, which implies that it is much probably a field star.
Moreover, in the case of young clusters, the $V$ vs $(U-B)$
CMD is better suited to separate members from non members.
The right panel of Fig.~4 indeed shows quite a tight MS,
and some stars rightward, well detached from the MS, which presumably
are all field stars.

   \begin{figure}
   \centering
   \resizebox{\hsize}{!}{\includegraphics{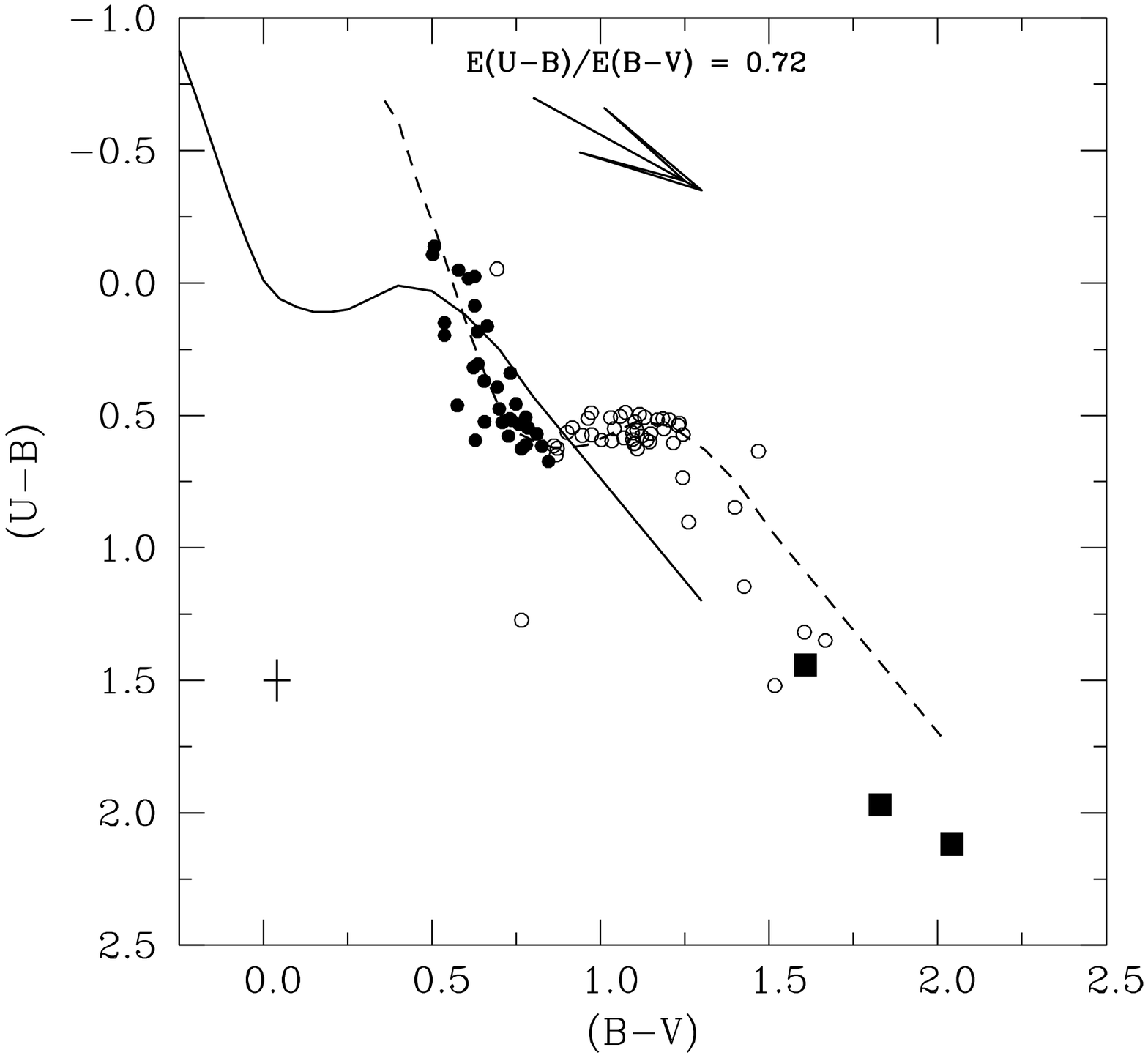}} 
   \caption{ Colour-colour diagram for all the stars in the field
of NGC~1220 having $UBV$ photometry. 
The solid line is the  Schmidt-Kaler (1982) empirical ZAMS,
whereas the dashed line is  the same ZAMS, but shifted by 
E$(B-V)$~=~0.70. Filled symbols indicate stars having
reddening in the range $0.60 \leq E(B-V) \leq 0.80$,
for which an unambiguous reddening solution has been possible.
The filled squares indicated the three bright very red stars
in Fig.~4, and the cross the typical colours error bar.}
    \end{figure}

\section{Two-colour diagram and members selection}
We derive cluster membership by grouping stars according
to their mean reddening. Individual reddening values have been
computed by means of the usual reddening free parameter
$Q$:

\[
Q = (U-B) - 0.72  \times (B-V)               ,
\]

\noindent
and the distribution of the stars
in the two colour Diagram, following the procedure outlined in detail
in Carraro (2002a,b), where the young open clusters Trumpler~15
and NGC~133 have been studied, respectively.
This method is a powerful one to isolate early 
spectral type 
(from $O$ to $A5$) stars 
having common reddening, which are most probably likely
cluster members (see, for a reference, the study of Trumpler~14
by Vazquez et al 1996). 
Moreover the reddening based members selection nicely
compares with -for instance-
proper motion based members selection (see Cudworth et al. 1993 and
Patat \& Carraro 2001 for some clusters in the Carina region).\\
Our results are shown in Fig~5,
where we plot all the stars having $UBV$ photometry
in the two-colour Diagram. The solid line is an empirical
ZAMS from Schmidt-Kaler (1982).
In this figure, we have plotted with filled symbols all
the stars having a mean reddening E$(B-V) = 0.70 \pm 0.15$ mag
(31 stars in total) previously determined, which obviously crowds
close to a ZAMS shifted by E$(B-V)=0.70$ mag (dashed line).
All these stars have common reddening and spectral type earlier than
$A5$. The spread in reddening seems to suggest that some
differential reddening is present across the cluster surface.
This is not unexpected, due to the cluster position in the Galactic 
thin disk.\\

The other stars are plotted with open symbols. For these latter
stars an unambiguous reddening solution is not possible, since they are 
located in a region  where larger reddening ZAMS cross the E$(B-V)$ =0.70
mag ZAMS, rendering not possible to effectively
disentangle members from non members. Three stars are plotted with filled
squares. They are the three very red stars in Fig.~4. From their
position in the two-colour diagram we argue that these stars are late 
spectral type ($K$ to $M$) low reddening, nearby stars.\\

Finally, more information can be derived by considering the distribution of 
the stars in the reddening corrected CMD (see Fig.~6). In this figure we have
plotted all the likely early spectral type members.
To guide the eye, two ZAMS have been drawn. The solid one, which fits 
the distribution of the bulk of member stars, has been shifted by 
$(m-M)_o~=~11.3\pm0.2$ (see also next section).
The dotted one has been placed to mimic the expected position
of presumed unresolved binary stars. One can readily see that most
of member stars fall close to the ZAMS location.
There are actually some exceptions (stars \#1, \#3, \#9, \#14, 
and \#67), which are clearly off (too red or too blue) the MS.
We are not going to consider these stars as NGC\,1220
 likely members.\\
In conclusion, we would like to argue that the population
of stars having E$(B-V) = 0.70 \pm 0.15$ mag (26 stars)
identifies the brightest members of the open cluster
NGC\,1220. Later spectral type stars are more difficult
to be detected, due to the degeneracy described above.
The final photometry of the likely candidate members
of NGC\,1220 is listed in Table~3, 
together with individual reddenings and photometric spectral
types. The latter have been obtained by deriving intrinsic colours
from observed colours and reddenings. We use Johnson (1966)
intrinsic colours to infer approximate spectral types.
The typical uncertainty in the spectral type is $\pm$2,
due to both photometric errors and the difficulty
to estimate the luminosity classes.

   \begin{figure}
   \centering
   \resizebox{\hsize}{!}{\includegraphics{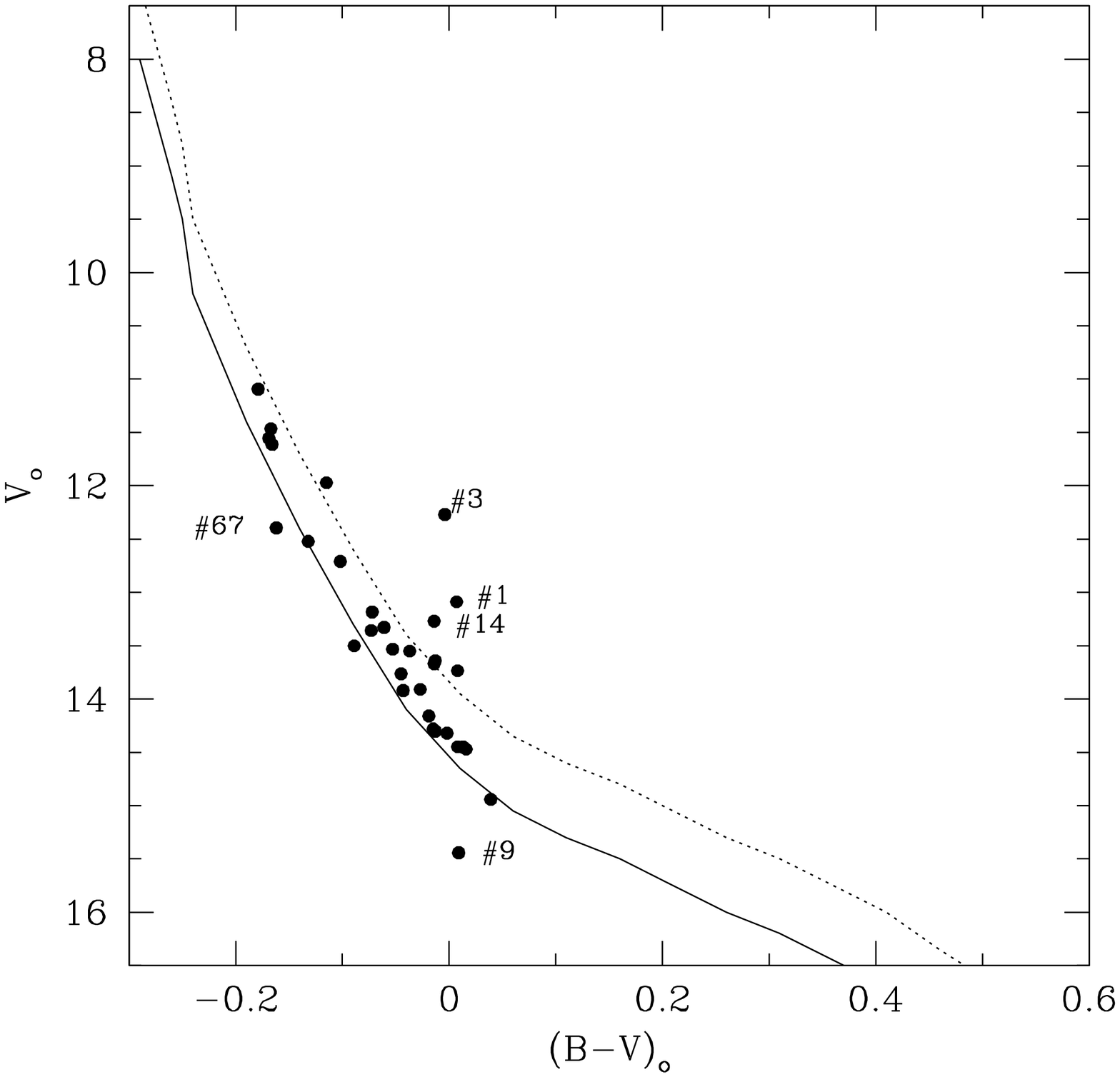}} 
   \caption{ Distribution of early spectral type candidate members 
in the reddening corrected  CMD. The solid line is an empirical ZAMS
shifted by $(m-M)_V~=~13.5\pm0.2$ mag, whereas the dotted line is the same ZAMS, but
0.70 mag brighter, drawn to mimic the location of the unresolved
binary stars.}
    \end{figure}

   \begin{figure}
   \centering
   \resizebox{\hsize}{!}{\includegraphics{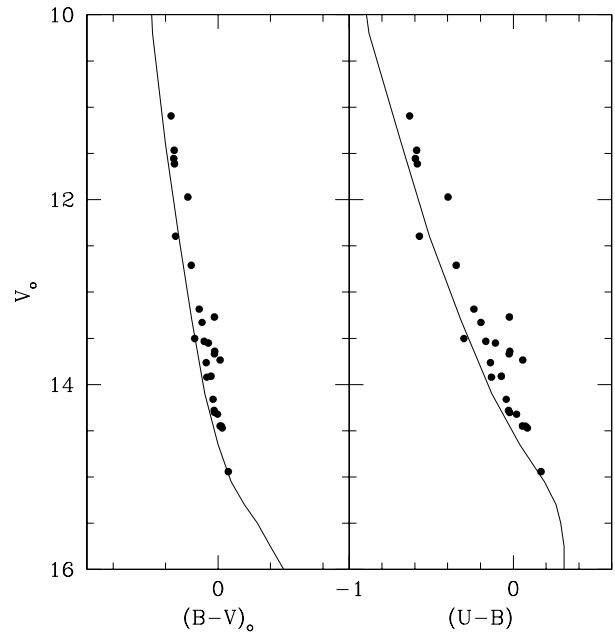}} 
   \caption{ Reddening corrected CMDs of the likely member
  stars in the region of NGC~1220. The solid line is an empirical
ZAMS shifted  by $(m-M)_V~=~13.50$ mag.}
    \end{figure}

\begin{table*}
\tabcolsep 0.40cm
\caption{Photometry of likely 
member stars in the field of NGC~1220. In the last column,
{\it p.n.m.} means probable non member.}
\begin{tabular}{cccccccccccc}
\hline
\hline
\multicolumn{1}{c}{ID} &
\multicolumn{1}{c}{$X (pixel)$}&
\multicolumn{1}{c}{$Y(pixel)$}&
\multicolumn{1}{c}{$V$}  &
\multicolumn{1}{c}{$(B-V)$} &
\multicolumn{1}{c}{$(U-B)$}  &
\multicolumn{1}{c}{$Q$} &
\multicolumn{1}{c}{E$(B-V)$} &
\multicolumn{1}{c}{{\it Sp.Type}} &
\multicolumn{1}{c}{$Notes$} \\
\hline
   1&   967.330&    15.690&    14.848&     0.575&     0.462&  0.048&    0.568& A1& p.n.m\\
   3&   228.090&   285.270&    14.480&     0.709&     0.526&  0.016&    0.713& B9& p.n.m.\\
   4&   567.300&   397.170&    13.539&     0.502&    -0.107& -0.468&    0.669& B7& \\
   5&   475.760&   433.540&    14.387&     0.664&     0.163& -0.315&    0.779& B6& \\
   6&   520.020&   483.510&    13.923&     0.579&    -0.049& -0.466&    0.745& A4& \\
   9&   492.600&     9.870&    17.667&     0.727&     0.578&  0.055&    0.718& A2& p.n.m.\\
  13&   688.280&   235.520&    15.743&     0.656&     0.524&  0.052&    0.648& A2& \\
  14&   851.920&   237.320&    15.247&     0.537&     0.198& -0.189&    0.610& B8& p.n.m.\\
  15&   536.950&   276.220&    15.442&     0.537&     0.150& -0.237&    0.626& B7& \\
  16&   864.100&   327.730&    16.795&     0.766&     0.626&  0.074&    0.750& A3& \\
  17&   449.040&   337.210&    16.838&     0.779&     0.611&  0.050&    0.771& B9& \\
  23&   570.490&   419.710&    15.627&     0.623&     0.319& -0.130&    0.676& B9& \\
  26&   550.960&   473.700&    15.491&     0.637&     0.305& -0.159&    0.698& B9& \\
  30&   690.850&   503.320&    13.219&     0.507&    -0.138& -0.503&    0.686& B5& \\
  35&   520.440&   538.900&    15.966&     0.737&     0.519& -0.012&    0.750& B9& \\
  36&   373.510&   541.340&    16.054&     0.694&     0.393& -0.107&    0.739& B9& \\
  39&   716.750&   577.490&    16.697&     0.759&     0.534& -0.012&    0.772& A0& \\
  48&   126.610&   791.540&    14.783&     0.608&    -0.017& -0.455&    0.770& B5& \\
  58&   246.520&   214.460&    16.085&     0.655&     0.370& -0.102&    0.698& B9& \\
  64&   494.780&   423.660&    15.987&     0.749&     0.457& -0.082&    0.786& B9& \\
  66&   422.610&   460.180&    14.023&     0.627&    -0.024& -0.475&    0.796& B5& p.n.m.\\
  67&   440.650&   525.220&    14.875&     0.627&     0.086& -0.365&    0.759& B5& \\
  68&   556.110&   530.450&    14.997&     0.636&     0.183& -0.275&    0.738& B6& \\
  69&   605.160&   540.530&    16.387&     0.700&     0.475& -0.029&    0.719& A0& \\
  78&   537.330&   350.389&    17.033&     0.846&     0.674&  0.065&    0.833& A3& \\
  79&   528.560&   378.300&    15.982&     0.732&     0.512& -0.015&    0.746& B9& \\
\hline
\end{tabular}
\end{table*}

\subsection{Distance and age}
In Fig.~7 we plot the reddening corrected CMDs for the likely
member stars above determined. 
In both diagrams we have overimposed the empirical
Schmidt-Kaler (1982) ZAMS, shifted by $(m-M)_o =11.3\pm0.3$ mag,
which provides a nice fit of the stars distribution.
This implies that NGC~1220 is located $1800\pm200$ pc away from the Sun,
where the uncertainty mirrors the difficulty of the fit
due to the almost vertical structure of the MS.\\
From the location of the stars in the $(B-V)$ vs. $(U-B)$ plane,
we infer that the stars spectral types range from $B5$ to
$A5$ by deriving the absolute colours from the ZAMS at the same 
position of the stars (see also Table~3 and previous section). 
If the stars having $B5$ spectral type are still along the MS,
we derive an age around 50 Myrs for 
NGC~1220 (Girardi et al. 2000).\\

\noindent
We checked this age estimate by fitting CMDs of NGC\,1220 with 
theoretical isochrones from Padova models 
(Girardi et al. 2000). The result is presented in Fig.~8.
In the left panel we plot the NGC\,1220 stars in the plane
$V$ vs $(B-V)$, and overimposed two solar metallicity isochrones.
The solid line is for the age of 60 Myr, the dashed one for the age of 250 Myr.
This latter has been drawn with the intention to try to fit
the three very red stars previously mentioned. The same isochrones
have been overlaid to NGC\,1220 stars in the $V$ vs $(U-B)$
diagram (right panel).
The overall fit is very good in both diagrams. However,
by closely inspecting only the left panel, one could not
definitely rule out the larger age for NGC\,1220, since
the dashed isochrone (250 Myr) nicely fits the bulk of the stars
but for only the two bright stars above the TO, which nevertheless
might be accounted for by invoking their possible binary nature.\\
In this respect the CMD in the right panel ($V$ vs $(U-B)$)
helps a lot in solving the mystery. In fact in this colour combination
the colour separation of the bluest stars is much wider. It results very 
clearly that only the 60 Myr isochrone provides a good fit to the MS stars,
thus ensuring us that all the stars red-wards the MS are simply field stars.
Therefore we conclude that NGC\,1220 is a young open cluster,
about 60 Myr old.

   \begin{figure*}
   \centering
   \resizebox{\hsize}{!}{\includegraphics{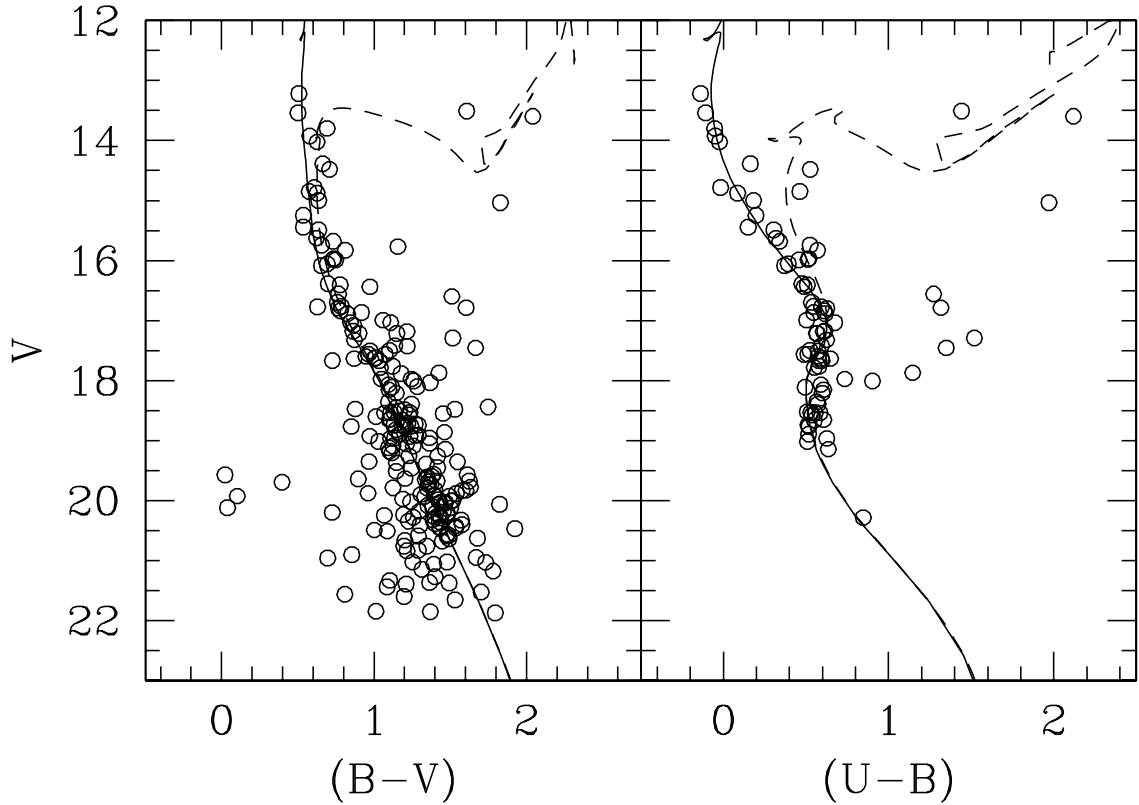}} 
   \caption{CMDs of all the 
  stars in the region of NGC~1220. The solid line is a solar 
metallicity 60 Myr
isochrone from Girardi et al. (2000), whereas the dashed one is 
a 250 Myrs isochrone.}
    \end{figure*}

The Galactocentric coordinates are X = -9.43 (X $<$0 means our side of the
Galaxy), Y = 1.08
and Z = -0.12 kpc, considering the distance to the Galactic center
to be R = 8 kpc (Reid 1993). The cluster is thus outside the
solar circle, in the Perseus arm (Taylor \& Cordes 1993).
However,
we notice that NGC\`1220 is fairly high
above the Galactic plane with respect to his age. 
In fact, it takes at least 10$^7$ yr for a young cluster
with a typical velocity of 10 km s$^{-1}$ to move about 100 pc.\\

\noindent
Combining together the estimated age, distance and position in the
Galaxy, we conclude that NGC\,1220 is a genuine Galactic 
this disk star cluster, presently located relatively high
above the Galactic plane, but presumably formed within the thin disk.

\section{Conclusions}
In this paper we have presented new CCD UBV photometry
for the stars in the field of NGC~1220, and provide
the first estimates of its fundamental parameters.\\
Our findings can be summarized as follows:

\begin{itemize}
\item NGC~1220 is a compact 20-25 stars group, with a radius of 
1.5-2.0 arcmin, which turns into 0.79-1.05 pc
at the distance of the cluster;
\item we identified 26 likely members with spectral type earlier than
$A5$ on the basis of the reddening and the position in the reddening
corrected CMDs;
\item the cluster turns out to be located about 1800 pc away from the 
Sun in the Perseus spiral arm;
\item we estimate a reddening $E(B-V)=0.70\pm0.15$;
\item the probable age is     around  60 Myrs;
\item NGC\,1220 is presently located 120 pc above the Galactic
plane, relatively high with respect to its age. With the available data
it is not possible to conclude whether the cluster formed 
at some distance above the galactic plane,
or formed well within the thin disk and then moved away. This latter
hypothesis would imply  a non negligible vertical motion of the cluster.
\end{itemize}

\noindent
We would finally like to note that more precise estimates of the cluster 
age can be derived by obtaining spectroscopic classification of the 
brightest stars. Moreover a proper motion study would permit
to better distinguish NGC\,1220 physical members.

\begin{acknowledgements}
This study made use of Simbad and WEBDA catalogs.
We acknowledge partial financial support from
the brazilian agencies CNPq and Fapesp,
and Ministero dell'Universit\`a e della Ricerca
Scientifica e Tecnologica (MURST) under the program
on 'Dynamics and Stellar Evolution in Globular Clusters:
a Challenge for New Astronomical Instruments'.
 
\end{acknowledgements}

\end{document}